# LINEAR COLLIDER PHYSICS[*]

DAVID J. MILLER

*Physics and Astronomy, University College London, Gower Street, London WC1E 6BT, UK*

The International Linear Collider has a rich physics programme, whatever lies beyond the standard model. Accurate measurement of the top quark mass is needed to constrain the model or its extensions. If there is a light Higgs boson the LHC should see it, but the ILC will pin down its characteristics and test them against model predictions. If Supersymmetric particles appear the ILC will measure a complementary set of them to those seen at the LHC, and may allow extrapolation to the Grand Unified scale. And if a strong electroweak sector is indicated the ILC will be sensitive to the presence of new structures in difermion and diboson systems up to higher masses than the direct search range of the LHC. Beyond the LHC and ILC there could be need for a multi TeV lepton collider.

## 1. Context

### 1.1. *What physics needs*

The Large Hadron Collider is needed as soon as possible, for many good reasons[1]. We also need two LCs: the first, also as soon as possible, will be the International Linear Collider (ILC) which has made great practical progress at this conference; then, in the longer term, a multi-TeV lepton collider which will probably use the CLIC dual-beam technology[2]. The case may also develop for a higher energy hadron collider.

### 1.2. *Choice of technology*

This week's news is that the ILC will be based on superconducting radiofrequency cavities, as recommended by the International Technology Recommendation Panel (ITRP)[3]. The brief to the ITRP included the following definition of the scope and parameters which will be required for the machine to meet its physics goals, as currently understood.

#### Scope and parameters (summary)[4]

- **Baseline.**
$\sqrt{s}$ = 200 to 500 GeV;
Integrated $\mathcal{L}$uminosity 500fb$^{-1}$ in 1$^{st}$ 4 yrs;
80% electron polarisation
2 interaction regions with easy switching.

- **Upgrade**.
$\sqrt{s}$=1 TeV, with $\int \mathcal{L}dt$ = 1 ab$^{-1}$ in 4 years.

- **Options**.
e$^-$e$^-$ collisions
50% positron polarisation
"Giga-Z"= high $\mathcal{L}$ at Z and WW threshold;
Laser backscatter for $\gamma\gamma$ and e$^-\gamma$ collisions;
Doubled $\mathcal{L}$ at 500 GeV;

The baseline is the essential minimum for which funding will initially be requested. But this is not a "sub TeV collider" as some persist in calling it. The upgrade is an integral part of the programme. ITRP has actually pressed for a machine design which could give more than 1 TeV if technical developments allow. The baseline design should not include features which rule out eventual addition of any of the options. Choice between the options will depend upon the needs of physics. It is possible that the e$^-$e$^-$ and polarised positron options could be added to the baseline if R&D shows they can be included at low cost.

### 1.3. *Physics cases for future machines*

Particle physics and cosmology both predict strong new signals at a scale of 1 TeV. On this basis the case for the LHC as a discovery

---

machine has been made and accepted. It will look into a large part of the region where new physics should be.

On the same basis the case for the ILC has also been made in regional and worldwide workshops over many years, gathered together in the TESLA TDR[5], the Snowmass Resource Book[6] and the GLC project report[7]. Over 2600 physicists have signed a consensus document[8] in support of it. The significance of the linear collider's contribution strengthens with each update of the report[9] from the LHC/LC Study Group. This talk presents some highlights of the case.

The physics cases for CLIC, for a muon collider or for a very large hadron collider would follow from the results at the LHC and ILC.

## 2. Programme for the ILC

### 2.1. *Roadmap*

There are many different predictions for the kind of new physics which will emerge at the TeV scale, but in every scenario considered the ILC will make an essential contribution. How the programme develops, and which options are chosen, will depend upon where we find ourselves in Figure 1.

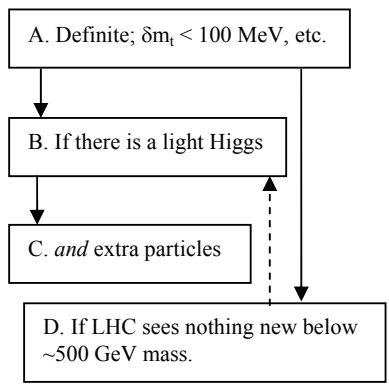

Figure 1. Roadmap for the ILC

### 2.2. *Top quark physics, the definite programme.*

At the startup of the ILC we will go to box A. LEP has already demonstrated how the parameters of the Standard Model (SM) are tied down by detailed precision analysis of $e^+e^-$ to 2 fermions and to 4 fermions, including the WW and to ZZ channels. At 500 GeV, with much more luminosity, the ILC will greatly tighten these constraints. But the most urgent task will be in the 6-fermion channel, determining the parameters of the top quark – particularly its mass.

The crucial role of the top quark in the structure of the Standard Model is illustrated by the effect of the recent result from the D0 experiment which changes the PDG average of the top quark mass from 174.3±5.1 GeV to 178.0±4.3 GeV, a shift of ~1σ. The resultant change in the predicted mass of the Higgs boson is +20 GeV, bringing the $\chi^2$ minimum of the LEP E-W working group's "blue band" plot[10] to ~114 GeV; close to the minimum allowed mass from the LEP SM Higgs search.

The baseline programme of the ILC includes the promise to measure the top quark mass to a precision of better than ±100 MeV, an order of magnitude better than can be done at the LHC. A recent paper has shown[11] that if the precision on the top quark mass is significantly worse than this it becomes the dominant error among the whole set of SM parameters, and it has a similar dominant effect on fits to SUSY models; see Figure 2, for an example in the Minimal Supersymmetric Standard Model (MSSM).

If the top mass is measured in a threshold scan at the ILC, QCD scheme-dependence[12] contributes about ±40 MeV to the error. To match this we need to measure the absolute energies of the linac beams to ~1 part in $10^4$ and to use the acollinearity of Bhabha scattering events to monitor the effect of beamstrahlung on the luminosity spectrum. Recent studies[13] show that systematic effects should be manageable, see Figure 3. R&D is needed to optimise methods.

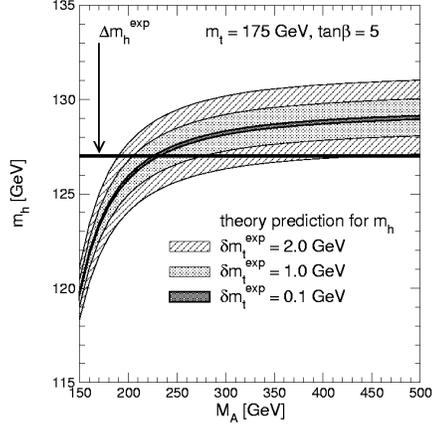

Figure 2. Sensitivity of an MSSM model[11] to the precision on $m_t$. A is the CP-odd Higgs boson, h is the lightest Higgs. The LC value of $\Delta m_h$ is assumed (see below).

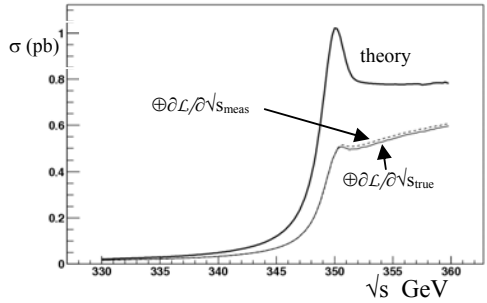

Figure 3. Modification of the top-antitop threshold for a Monte Carlo sample when either the true or a "measured" luminosity spectrum is assumed[14]. The top mass fitted to the data shifts by only 48 MeV, even before correction for known measurement bias.

### 2.3. *If there is a light Higgs boson.*

The direct, expected, route to box B of the roadmap is by discovery of a Higgs-like object at the LHC. But ingenious theorists[15] have also managed to devise light Higgs models which would be invisible at LHC but seen at ILC, giving an alternative route via the dotted upward arrow from box D to box B in Figure 1.

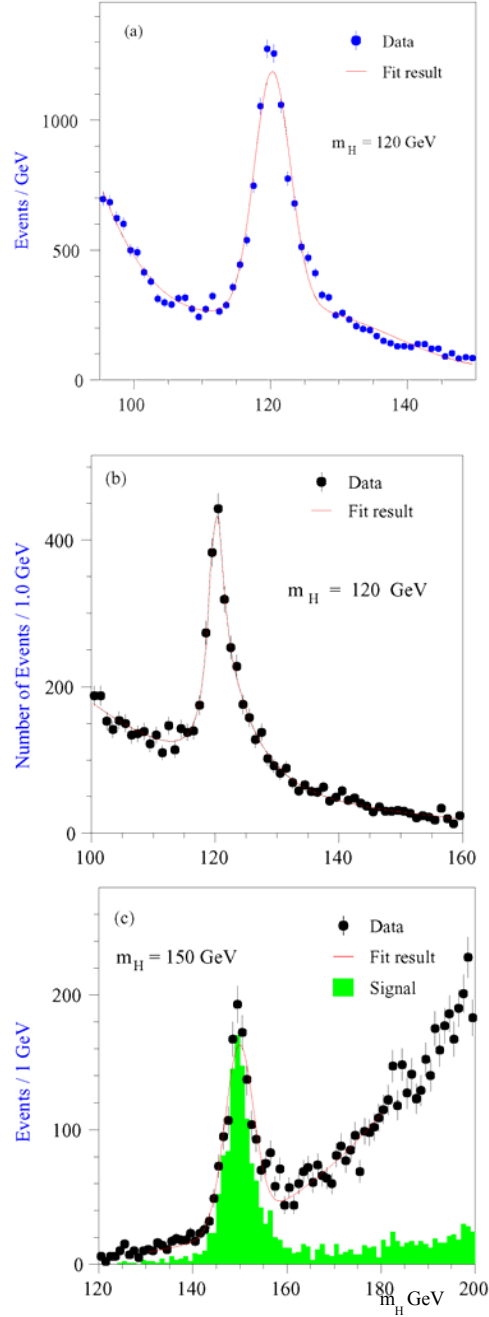

Figure 4. Fits[5] to light Higgs boson mass peaks; 500 pb⁻¹ at √s=350 GeV. (a) $ZH\rightarrow q\overline{q}b\overline{b}$, (b) $ZH\rightarrow q\overline{q}ll$, (c) $ZH\rightarrow q\overline{q}WW$.

If there is a light Higgs boson the ILC will see it and answer important questions about it:

- does it have SM couplings?
- what is its precise mass?
- does it have light scalar partners?

Many scenarios are being investigated, in the LHC/LC study[9] and elsewhere[16].

The quality of data which will be available for determining the Higgs boson mass can be seen in Figure 4. Recent studies with different detector simulations and realistic beam-quality assumptions have confirmed that the resolutions shown in table I are good estimates.

Table I. Resolution on assumed Higgs masses using constrained fits[19]; 500 pb⁻¹ at √s=350 GeV.

| $m_H$(GeV) | Channel | $\delta m_H$(MeV) |
|---|---|---|
| 120 | $q\bar{q}l\bar{l}$ | ±70 |
| 120 | $q\bar{q}b\bar{b}$ | ±50 |
| 120 | Combined | ±40 |
| 150 | $l\bar{l}$ +recoi | ±90 |
| 150 | $q\bar{q}WW$ | ±130 |
| 150 | Combined | ±70 |
| 180 | $l\bar{l}$ +recoi | ±100 |
| 180 | $q\bar{q}WW$ | ±150 |
| 180 | Combined | ±80 |

With $m_H$<200GeV there will be copious production of $t\bar{t}H$ in LHC final states, giving ~15-20% determination of the coupling[20] when Higgs branching ratios from

Table II. Comparison[5] of expected accuracies on SM-like Higgs parameters, including couplings and their ratios, at LHC (2×300fb⁻¹) and ILC (500fb⁻¹). (* New results on top-quark Yukawa coupling and Higgs self-coupling are discussed in the text.)

| | $m_H$ (GeV) | $\delta X/X$ LHC | $\delta X/X$ ILC |
|---|---|---|---|
| $m_H$ | 120 | 9×10⁻⁴ | 3×10⁻⁴ |
| $m_H$ | 160 | 10⁻³ | 3×10⁻⁴ |
| $\Gamma_{tot}$ | 120-140 | - | .04 to .06 |
| $g_{Hw\bar{w}}$ | 120-140 | - | .02 to .04 |
| $g_{Hd\bar{d}}$ | 120-140 | - | .01 to .02 |
| $g_{HWW}$ | 120-140 | - | .01 to .03 |
| $g_{Hw\bar{w}}/g_{Hd\bar{d}}$ | 120-140 | - | .023 to .052 |
| $g_{Hb\bar{b}}/g_{HWW}$ | 120-140 | - | .012 to .022 |
| $g_{Ht\bar{t}}/g_{HWW}$ | 120 | *see | *text |
| $g_{HZZ}/g_{HWW}$ | 160 | 0.050 | 0.022 |
| *CP* test | 120 | - | 0.03 |
| $\lambda_{HHH}$ | 120 | *see | *text |

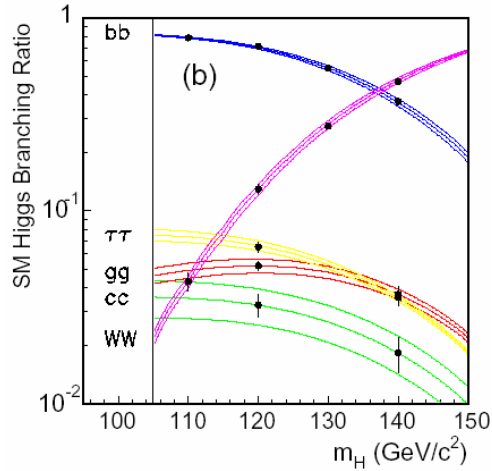

Figure 5. Predicted SM Higgs branching ratios[17] Error bars show simulated experimental accuracy. Bands give theoretical uncertainties.

The estimated ILC errors on SM Higgs boson branching ratios are shown in Figure 5. From table II it can be seen just how comprehensively the ILC will be able to characterize the properties of a light Higgs boson with SM-like couplings. Newer results on LHC[18] and ILC precision do not significantly change the comparison – except for $g_{Ht\bar{t}}$, the top-quark Yukawa coupling.

the 500 GeV ILC are combined with the LHC data (upper curve in Figure 6). The region around $m_H$=120GeV benefits particularly from the ILC's precision on $H \to b\bar{b}$. The lower grey solid curve in figure 6 shows the effect on the combined analysis of upgrading the energy of the ILC in order to use $e^+e^- \rightarrow t\bar{t}H$, which rises from a negligible cross section at $\sqrt{s}$=500GeV to a maximum at 800GeV for $m_H$=120GeV. The predicted precision on $g_{Ht\bar{t}}$ at this mass would be 5.5%, and less than 14% all the way up to $m_H$=200GeV.

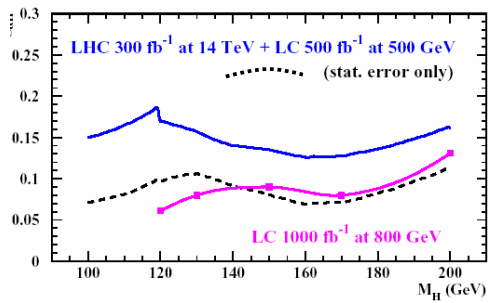

Figure 6. Relative precision on top quark Yukawa coupling[21]. Upper solid curve is LHC + 500 GeV ILC (dotted is corresponding statistical error). Lower grey curve adds 800 GeV ILC.

Earlier studies of the precision on the Higgs boson self coupling $\lambda_{HHH}$ are now being revised[22], both for LHC and ILC. If $m_H{\sim}120$ GeV there are prospects of a decent measurement at ILC: ~23% for $\sqrt{s}$=500GeV[23], and a recent preliminary result[24] of ~10% for $\sqrt{s}$=1TeV. If $m_H{\sim}140$GeV the ILC ceases to be as sensitive as the LHC, but even there it will not be easy to establish to better than 95% confidence that the coupling is nonzero[25].

The "CP test" entry on table II indicates[26] that a 3% CP-violating contribution to the ZH cross section would be detectable at the 1-σ level; for 500 fb$^{-1}$ at $\sqrt{s}$=350GeV.

## 2.4. *Precision can bring discoveries*

Throughout the roadmap of Figure 1, especially if there is a light Higgs boson, the ILC will achieve much higher precision than LHC - by more than an order of magnitude in some measurements, resolving the unresolvable in others. The power of extra precision is illustrated by recent results on the angular moment analysis of the cosmic microwave background, where the WMAP[27] results have revealed sensitivity to cosmic parameters which were inaccessible to COBE. For instance, in fitting the WMAP data it is now necessary to include a dark energy term as well as dark matter. Further extra precision from the Planck satellite is eagerly awaited by cosmologists; especially the information which polarization will bring. Analogies to the ILC should not be forced too far, but they are visible.

## 2.5. *New particles with a Higgs boson.*

Arriving in box C on the roadmap of Figure 1 probably means that the mechanism for electroweak symmetry breaking is emerging. Many scenarios are being investigated; in particular supersymmetric (SUSY) models, almost all requiring a light Higgs boson that would be thoroughly characterised by the ILC in an extension of the studies described above.

The lightest neutralino $\tilde{\chi}_1^0$, in the MSSM and in many other versions of SUSY, is particularly interesting since it is a candidate for the main constituent of cosmological Cold Dark Matter. Even if the only open SUSY channel is $e^+e^- \rightarrow \tilde{\chi}_1^0 \tilde{\chi}_2^0$ it can be clearly recognized at the ILC, above standard model backgrounds, and the two neutralino masses would be measured with good resolution[28]. Constraints from cosmology[29] have been applied by a number of groups to sets of SUSY models. Figure 7, from the latest of these exercises[30], gives a better representation of the reach of the ILC than the plot shown in the final talk[31] of this conference which ignored the $e^+e^- \rightarrow \tilde{\chi}_1^0 \tilde{\chi}_2^0$ channel. A large fraction of the allowed combinations of parameters scanned for Figure 7 would give signals accessible to ILC, at least through $e^+e^- \rightarrow \tilde{\chi}_1^0 \tilde{\chi}_2^0$. The isolated spot at low neutralino masses in Figure 7 is from the "focus point" region of parameter space (also omitted from ref[31]), in parts of which the discovery reach of the ILC exceeds that of LHC[32].

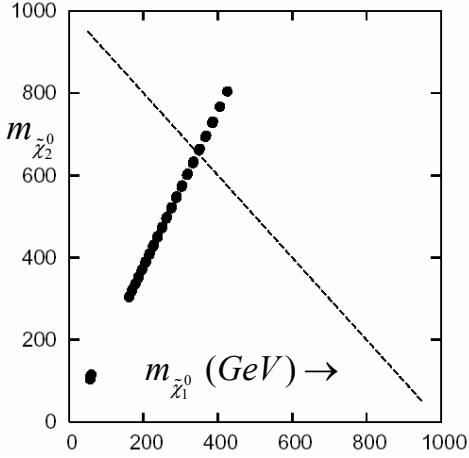

Figure 7. Masses of $\tilde{\chi}_1^0$ and $\tilde{\chi}_2^0$ from a search[30] of MSSM models satisfying cosmological constraints[29], with $\tan \beta = 10, 20, 35$ and A=0. The dotted line shows the reach of the 1TeV ILC.

Another survey of sets of models, reported in ref[33], shows that the combination of LHC+ILC would cover virtually the same set of possible new SUSY particles as would a 3 TeV CLIC.

In the LHC/LC report[9] there has been a thorough study of the interplay between ILC and LHC information for a particular allowed point SPS1a[34] in the parameter space of the MSSM. Some results are shown in table III. Note the much better precision of the ILC on the masses of the lightest Higgs, of the light neutralinos (by two orders of magnitude for $\tilde{\chi}_1^0$) and of the sleptons. LHC input is essential for most of the coloured objects, i.e. for the squarks and the gluino. The combination of the two colliders is needed to get the best masses for $\tilde{\chi}_2^0$, $\tilde{\chi}_4^0$. The precision of the ILC comes in some cases from threshold scans, as described in 2.2 above for the top mass, in other cases from precise measurements of sharp edges in spectra of final state particles, as shown in Figure 8.

Table III. Masses, and their predicted measurement errors (in GeV), at SUSY benchmark point SPS1a[34] for a representative sample of its MSSM spectrum. The last column is the result of a fit which includes input from both colliders[35].

| Particle | $m_{SPS1a}$ | LHC | ILC | LHC+ILC |
|---|---|---|---|---|
| h | 111.6 | 0.25 | 0.05 | 0.05 |
| A | 399.1 | | 1.5 | 1.5 |
| H | 399.6 | | 1.5 | 1.5 |
| $H^+$ | 407.1 | | 1.5 | 1.5 |
| $\tilde{\chi}_1^0$ | 97.03 | 4.8 | 0.05 | 0.05 |
| $\tilde{\chi}_2^0$ | 182.9 | 4.7 | 1.2 | 0.08 |
| $\tilde{\chi}_3^0$ | 349.2 | | 4.0 | 4.0 |
| $\tilde{\chi}_4^0$ | 370.3 | 5.1 | 4.0 | 2.3 |
| $\tilde{\chi}_1^\pm$ | 182.3 | | 0.55 | 0.55 |
| $\tilde{\chi}_2^\pm$ | 370.6 | | 3.0 | 3.0 |
| $\tilde{g}$ | 615.7 | 8.0 | | 6.5 |
| $\tilde{t}_1$ | 411.8 | | 2.0 | 2.0 |
| $\tilde{b}_1$ | 520.8 | 7.5 | | 5.7 |
| $\tilde{u}_1$ | 551.0 | 19.0 | | 16.0 |
| $\tilde{d}_1$ | 520.8 | 19.0 | | 16.0 |
| $\tilde{e}_1$ | 144.9 | 4.8 | 0.05 | 0.05 |
| $\tilde{\mu}_1$ | 144.9 | 4.8 | 0.2 | 0.2 |
| $\tilde{\tau}_1$ | 135.5 | 6.5 | 0.3 | 0.3 |
| $\tilde{\nu}_1$ | 188.2 | | 1.2 | 1.2 |

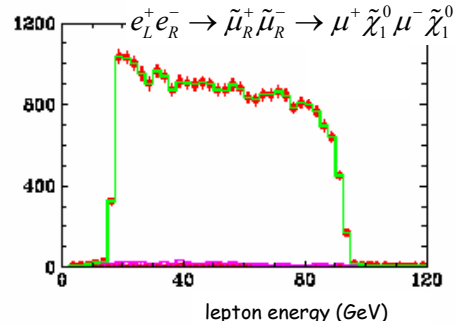

Figure 8. Simulated lepton spectrum[36], including backgrounds, for smuon pair production at point SPS1a with √s=400 GeV and $\mathcal{L}$=200 fb⁻¹.

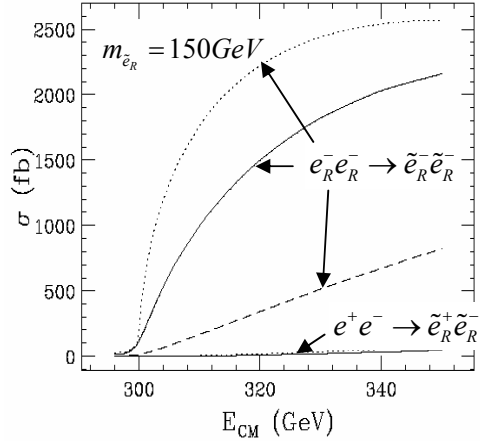

Figure 9. Threshold shapes for right handed selectron production[37]. The solid line corresponds to realistic beam conditions.

The $e^-e^-$ and $e^-\gamma$ options are specially useful for seeing right handed selectrons. They can be produced in the S-wave, giving much bigger cross sections than in $e^+e^-$ (see Figure 9), and standard model backgrounds are automatically suppressed.

In a fit[38] to simulated data in the "phenomenological MSSM", a complete solution was only possible with both LHC and ILC inputs. The strongest constraints on 4 out of the 5 fundamental SUSY parameters came from the ILC measurements.

The ultimate goal, after fitting as many as possible of the masses and couplings in a verified SUSY model, will be to extrapolate to the GUT scale to get at the fundamental structure of the theory. Figure 10 shows how this might work out, in a magnified view close to the unification point for a minimal Supergravity (mSUGRA) model[39]. Note that the outer limits, from LHC alone, would be almost consistent with concordance between the three extrapolated couplings, whereas the tighter bands obtained from LHC+ILC suggest a lower crossing point for two of them; new physics close to the GUT scale.

## 2.6 Nothing new below ~500 GeV mass

A light Higgs boson is the most straightforward explanation of what we know about Electroweak symmetry breaking. But Peskin[40] has reminded us that the simple quartic Higgs potential is just a formula, not a physical theory. There are familiar arguments within a standard model framework that the Higgs must be light, but in thinking about the phenomenology of

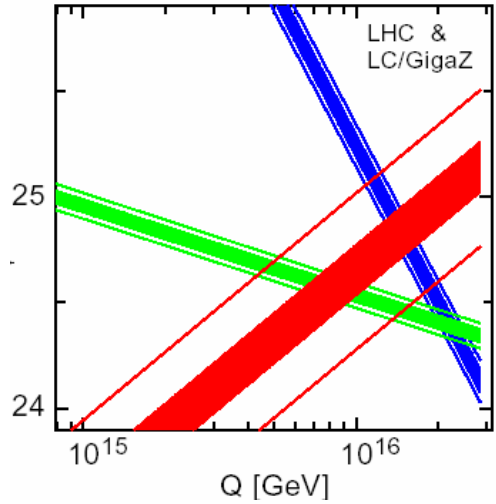

Figure 10. Renormalisation group extrapolation[39] to the region of the GUT scale of the three inverse gauge couplings in an mSUGRA model. Outer lines are with constraints from LHC; inner bands include ILC measurements.

extra dimensions theorists have devised "little Higgs models", "Heavy Higgs models", "Higgsless models" and more – many of them discussed in the LHC/LC report[9]. We could arrive at box D of the roadmap if one of them is true. Such models bring new s-channel structure above 1 TeV, coming either from extended gauge bosons or from new resonances. The LHC will look into this region, and should see some of the structure, if present, but it cannot cover the whole mass range. In fact, for some models, the indirect mass-reach of the ILC in the 2-fermion final state, even at $\sqrt{s}$=500GeV, goes beyond the ~5 TeV limit for direct discoveries at LHC . Figure 11 compares

LHC's reach with ILC's ultimate potential for Z´ discovery. With the 1 TeV upgrade and the Giga-Z option ILC can go beyond the LHC in all cases shown - even up to 19 TeV mass for the Kaluza Klein example on the right.

In the four fermion channel, study of angular distributions in WW final states allows the scattering amplitude to be extracted[41]. The ellipse at the left of Figure 12 represents a 95% confidence limit on this measurement if there were absolutely no strong WW interaction. The minimal effect of a strong interaction is represented by the LET (low energy theorem) point, which corresponds to an infinitely massive resonance. Other points are labeled with the mass of the WW resonance. The 500 GeV ILC can resolve a single resonance from the LET point if its mass is below 2.5 TeV, to be compared with a direct search limit of 1.9 TeV at the LHC. The 1 TeV ILC will be able to resolve a 4.1 TeV resonance from the LET point.

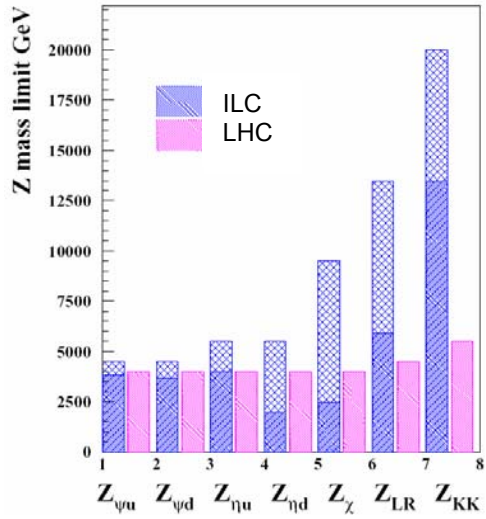

Figure 11. Sensitivity[42] for a new Z´ decaying to two fermions, in different models. The right hand box in each pair is for LHC. The denser part of each left hand box shows the range in which ILC would be able to establish a particle both by mixing and by interference.

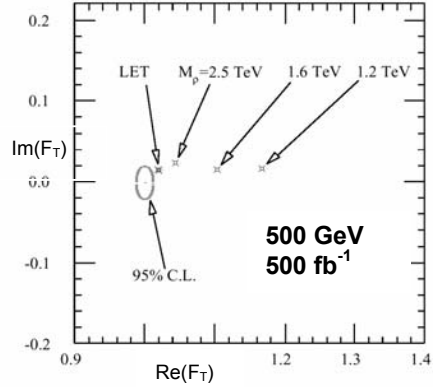

Figure 12. $W_L W_L$ scattering amplitude from $e^+ e^- \rightarrow WW$ [41].

Figure 13 shows how the properties of a vector resonance could be measured[43] at LHC, and how the measurement would be substantially improved by combination with ILC data.

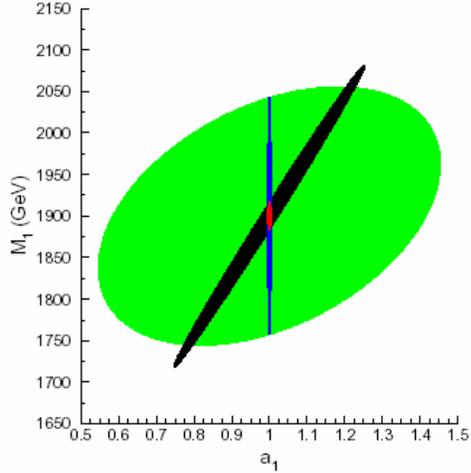

Figure 13. Measurement of the mass $M_1$ of a 1.9 TeV vector resonance in the WW channel at LHC alone (broad shaded ellipse) and with ILC data (narrow tilted ellipse). If the coupling to WW is set at the natural value ($a_1$=1) the errors on $M_1$ are shown by the long upright ellipse (LHC alone), or by the small central upright ellipse (with ILC).

### 3. Summary

Wherever we find ourselves on the roadmap of Figure 1 there is important work for the ILC to do. In box A it will place vital constraints on the standard model and on any extension, by making precise measurements of 2-fermion, 4-fermion and 6-fermion processes. Determining the top quark mass to better than 100 MeV is an important task which comes early in the baseline programme.

If we are in box B by the time ILC begins, the LHC will have seen the first signals for a light Higgs boson. The ILC will measure many more of its properties, with precision – and will look for deviations from the standard model, or from candidate extensions such as SUSY models.

In box C there is the prospect for the ILC to make precise measurements of the lighter members of the new particle spectrum from SUSY; especially of $\tilde{\chi}_1^0$ which would be well resolved in decay spectra. This – or another LSP – could be established as the relic particle which forms cosmological dark matter. If a coherent SUSY model were to emerge from the combined LHC and ILC data then masses and couplings from both colliders would be needed to constrain an extrapolation to the Grand Unified scale.

If LHC sees no new objects below 500 GeV mass before ILC begins, it is just possible that there is an invisible Higgs boson which ILC will reveal and study, but it is more likely that we have arrived in box D. New physics has to appear at the scale of a few TeV. The precise measurements of the ILC, augmented by the Giga-Z option, will be sensitive to new effects up to a higher scale than the direct reach of the LHC.

Whichever box we reach, there are likely to be loose ends which can only be studied at a higher energy machine, but both ILC and LHC data will be needed to point to whether the next step up in energy should be at a lepton or a hadron collider.

The cartoon summarises the summary.

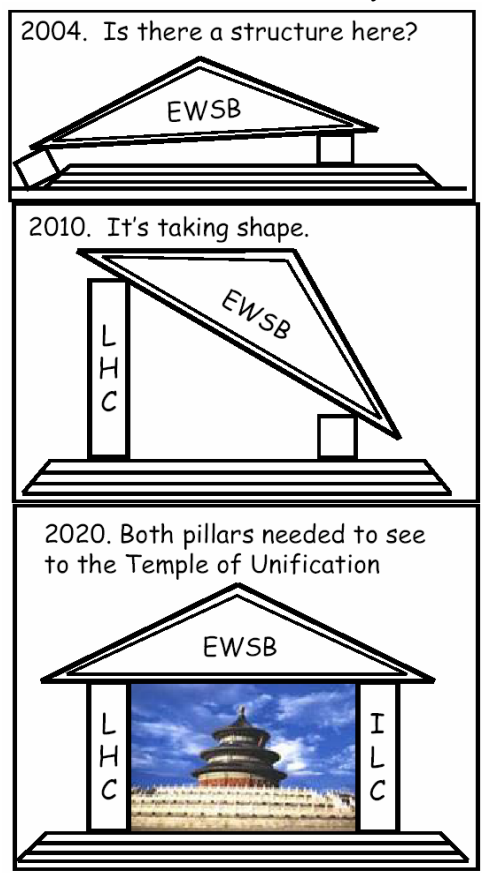